# The gravity field and interior structure of Dione

Marco Zannoni[1*], Douglas Hemingway[2,3], Luis Gomez Casajus[1], Paolo Tortora[1]

[1]Dipartimento di Ingegneria Industriale, Università di Bologna, Forlì, Italy

[2]Department of Earth & Planetary Science, University of California Berkeley, Berkeley, California, USA

[3]Department of Terrestrial Magnetism, Carnegie Institution for Science, Washington, DC, USA

[*]Corresponding author.

## Abstract

During its mission in the Saturn system, *Cassini* performed five close flybys of Dione. During three of them, radio tracking data were collected during the closest approach, allowing estimation of the full degree-2 gravity field by precise spacecraft orbit determination.

The gravity field of Dione is dominated by $J_2$ and $C_{22}$, for which our best estimates are $J_2 \times 10^6$ = 1496 ± 11 and $C_{22} \times 10^6$ = 364.8 ± 1.8 (unnormalized coefficients, 1-$\sigma$ uncertainty). Their ratio is $J_2/C_{22}$ = 4.102 ± 0.044, showing a significative departure (about 17-$\sigma$) from the theoretical value of 10/3, predicted for a relaxed body in slow, synchronous rotation around a planet. Therefore, it is not possible to retrieve the moment of inertia directly from the measured gravitational field.

The interior structure of Dione is investigated by a combined analysis of its gravity and topography, which exhibits an even larger deviation from hydrostatic equilibrium, suggesting some degree of compensation. The gravity of Dione is far from the expectation for an undifferentiated hydrostatic body, so we built a series of three-layer models, and considered both Airy and Pratt compensation mechanisms. The interpretation is non-unique, but Dione's excess topography may suggest some degree of Airy-type isostasy, meaning that the outer ice shell is underlain by a higher density, lower viscosity layer, such as a subsurface liquid water ocean. The data permit a broad range of possibilities, but the best fitting models tend towards large shell thicknesses and small ocean thicknesses.

## 1. Introduction

With a mean radius of 561 km, Dione is the fourth-largest moon of Saturn. It was discovered in 1684 by the Italian astronomer Giovanni Domenico Cassini, during observations made at the Paris Observatory. With a semi-major axis of approximately 6.26 Saturn radii ($R_S$) (377,400 km), Dione is in a 1:2 mean-motion resonance with the smaller moon Enceladus (252 km radius, 3.95 $R_S$ (237,900 km) semi-major axis). This resonance causes an orbital libration with a period of about 11 yr, and a circulation with a period of about 3.8 yr (Murray and Dermott, 1999). The resonance maintains a non-zero orbital eccentricity of both Enceladus and Dione, about 0.0047 and 0.0022, respectively. Dione has two co-orbital moons, Helene (17.6 km radius, discovered in 1980 from ground-based observations) and Polydeuces (1.3 km radius, discovered in 2004 from *Cassini* images), orbiting around the Lagrangian points L4 and L5, respectively.

Due to the small size and the large distance from the Earth, little was known about Dione's internal structure before the advent of the space era. From ground optical and spectroscopic observations, the surface was determined to be composed of almost pure ice. The mass of Dione was estimated by exploiting the mutual orbital perturbations with Enceladus, using ground-based astrometric observations (e.g. Kozai, 1976; Harper & Taylor 1993; Vienne & Duriez 1995; Dourneau & Baratchart 1999).

The knowledge about Dione advanced significantly with the Saturn flybys of the twin Voyager probes, in 1980 and 1981. Voyager 1 provided an almost global image coverage of the surface, revealing evidence of past geological activity, including resurfacing events and, possibly, cryovolcanic features (Smith et al., 1981; Plescia, 1983). The shape reconstructed from Voyager images, along with the mass given by Earth-based astrometric measurements, provided a first measure of the bulk density of Dione, at about 1440 kg/m$^3$ (Campbell & Anderson, 1989). This relatively high value was compatible with a mixture of about 55% by mass water ice (density 1000 kg/m$^3$) and 45% rock-metal (density 3000 kg/m$^3$). Subsequently, Jacobson (1995) obtained an improved estimation of the mass of Dione by measuring the orbital perturbations caused on Helene using ground-based astrometric and Voyager images. In preparation for the *Cassini* tour, Jacobson (2004) provided an updated estimate of the ephemerides and physical parameters of the Saturn system bodies, including the mass of Dione. An extensive data set was used, comprising Earth-based astrometry, Pioneer 11 and Voyagers' radiometric and optical data, and *Cassini* optical data acquired before the arrival in the Saturn system.

During its 13-year tour of the Saturn system, *Cassini* performed five close encounters of Dione, four of which were dedicated to the determination of its mass and gravity field, with the objective of constraining its internal structure. During the first two flybys, referred to as D1 (October 2005) and D2 (April 2010) according to the numbering scheme used by the *Cassini* project, radiometric data were collected only before and after (but not during) the closest approach (C/A), only allowing the estimation of the moon's mass. Using *Cassini* data acquired during the Saturn tour up to June 2006, including the data acquired during D1, Jacobson et al. (2006) provided an updated estimate of Saturn's gravity and pole orientation, and the masses of the satellites. In particular, the mass estimation of Dione improved by 1 order of magnitude, with the information coming mainly from astrometric and spacecraft imaging of Helene. The corresponding bulk density of Dione increased slightly to (1475.7 ± 3.6) kg/m$^3$. However, in the absence of measurements of the high-degree gravity harmonics, the internal structure could not be inferred. The first flyby dedicated to the determination of Dione's gravity field was D3 (December 2011). The analysis of the Doppler data acquired during the closest approach produced the first estimation of Dione's $J_2$ and $C_{22}$, suggesting that the moon is not compatible with the condition of hydrostatic equilibrium (Iess et al., 2012a). However, given the limited amount of data, the solution wasn't fully reliable. For this reason, during the extended mission, *Cassini* performed two other flybys of Dione with tracking during the closest approach, D4 (June 2015) and D5 (August 2015), to better characterize the moon's internal structure.

**Table 1: Summary of *Cassini* flybys of Dione. For each encounter, the table reports the date of *Cassini*'s closest approach (C/A), the minimum altitude reached, the orbital inclination, the Sun-Earth-Probe (SEP) angle, the number of Doppler points (60 s integration time), and the root mean square (RMS) of Doppler residuals at 60 s.**

| Flyby | Date of C/A | Altitude (km) | Rel. Vel (km/s) | Inclination (deg) | SEP (deg) | Doppler points | RMS at 60s (mm/s) |
|---|---|---|---|---|---|---|---|
| D1 | 11 Oct. 2005 | 495 | 9.1 | 120 | 69 | 829 | 0.035 |
| D2 | 07 Apr. 2010 | 503 | 8.3 | 0.5 | 163 | 410 | 0.027 |
| D3 | 12 Dec. 2011 | 98 | 8.7 | 175 | 53 | 1309 | 0.035 |
| D4 | 16 Jun. 2015 | 511 | 7.3 | 80 | 154 | 989 | 0.038 |
| D5 | 17 Aug. 2015 | 476 | 6.4 | 96 | 93 | 1144 | 0.035 |

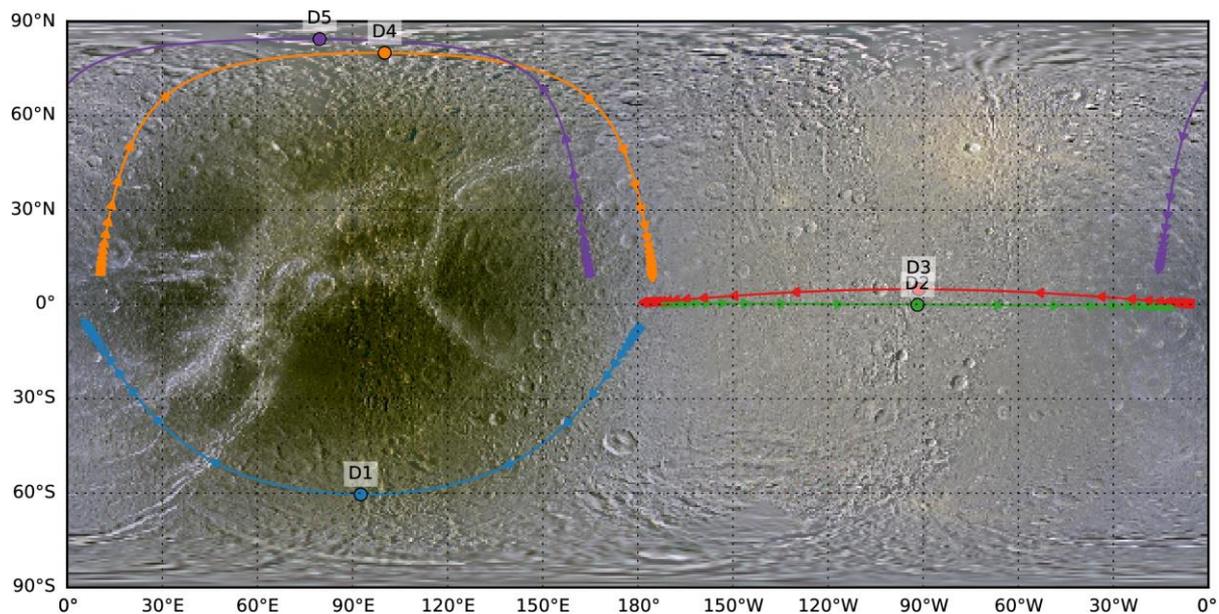

Figure 1: *Cassini* ground track on Dione during D1, D2, D3, D4 and D5, considering a time interval of ±15 minutes around the C/A (indicated by a circle). The ticks are separated by 60 s. The ground tracks are represented over a visible map of Dione produced by Paul Schenk (Lunar and Planetary Institute) from *Cassini* ISS data (NASA, JPL).

This paper presents the first estimation of Dione's quadrupole gravity field, obtained from the analysis of Doppler data acquired during all *Cassini* flybys of the moon. The main characteristics of Dione's *Cassini* flybys are summarized in Table 1, while the corresponding ground tracks are displayed in Figure 1. The flybys dedicated to gravity investigations, D3 and D5, provide good spatial coverage for the retrieval of the quadrupole gravity field. D3 was nearly equatorial, with an inclination at the C/A of about 175°. In order to de-correlate the estimation of $J_2$ and $C_{22}$, D5 was designed to be nearly polar, with an inclination at the C/A of about 96°. D3 also flew over the leading hemisphere, whereas D5 was over the trailing hemisphere. Being separated by only two months, D4 had an orbital geometry similar to D5. However, while D3 was characterized by a very low altitude at the C/A, less than 100 km, all the other flybys had a much higher altitude, of about 500 km, thus significantly reducing the sensitivity to the gravity field. The noise on X-band Doppler measurements is mainly due to the solar plasma and Earth troposphere (Asmar et al., 2005; Iess et al., 2012b). The former is correlated with the Sun-Earth-Probe (SEP) angle, which was larger than 50° during all encounters. The Doppler noise level around the C/A of the different flybys varies between a minimum of 0.021 mm/s, and a maximum of 0.036 mm/s.

This paper is organized as follows: Section 2 describes the data analysis approach for the estimation of Dione's gravity field, along with the spacecraft dynamical model, and the data selection and calibration procedure. Section 3 provides a geophysical interpretation of the results, by means of a combined analysis of Dione's estimated gravity and topography. Finally, Section 4 summarizes our findings and conclusions.

## 2. Gravity Analysis

### 2.1. Introduction

The gravity field of Dione was estimated by precisely reconstructing the trajectory of *Cassini* during the five close flybys of the moon. The estimation of the gravity field of Dione was based on the same procedure and techniques adopted in the previous gravity analyses of Saturn's moons performed by the *Cassini* Radio Science Team (Iess et al., 2012c; Iess et al., 2014; Tortora et al., 2016; Durante et al., 2019).

The main difference from past gravity analyses relates to the update of the ephemerides of Dione. Doppler data acquired around the pericenter of a flyby are very sensitive to the relative position of the *Cassini* spacecraft with respect to the moon. Outside the sphere of influence of Dione, which has a radius of about 2000 km, the data are sensitive mainly to the relative position of the spacecraft with respect to Saturn. During a close encounter, *Cassini* stays inside the sphere of influence of the moon for about 10 minutes. As a result, the orbit of Dione must be known at a level currently not met by the JPL satellite ephemerides, and so it must be estimated and updated as a part of the orbit determination procedure.

In previous work by Tortora et al. (2016) and others, the orbit of the moon under study was numerically integrated for the entire time span covered by the data, from an epoch prior to the first flyby, to after the last flyby. This approach ensures that the satellite trajectory is dynamically coherent.

However, as in the work of Durante et al. (2019) for Titan, in this work it was not possible to obtain a satisfactory fit of the data by estimating a single, coherent, orbit of Dione. This may be an indication of an incomplete dynamical model of the Saturn system, given the long timespan covered by *Cassini* data (D1 and D5 are separated by about 10 years), the poor sampling of the orbit with time (5 encounters), and the high level of accuracy of the data. Possible areas of improvement of the dynamical model are: the proper modelling of a time-variable or longitudinally-dependent component of the gravity of Saturn (Iess et al., 2019); the ephemerides of Enceladus, whose gravitational perturbations on Dione are relevant because of the orbital resonance; the evolution of Saturn's pole; the tidal interaction between Saturn and its moons. In particular, regarding the latter, recent measurements suggest that the tidal dissipation of Saturn is higher than predicted by standard tidal theories, and that it is not constant between the different satellites, as predicted by the resonance locking tidal theory (Lainey et al., 2012, 2017, 2020; Fuller et al., 2016). However, the accurate modelling of the motion of Dione inside the Saturn system was beyond the scope of this paper. Hence, the same approach followed by Durante et al. (2019) was adopted, estimating an updated orbit of the moon for each encounter. This over-parametrization causes an increase of the uncertainties of the quadrupole gravity coefficients up to 40%, but it ensures an unbiased estimate of Dione's gravity field.

### 2.2. Dynamical Model

The adopted dynamical model included all the relevant accelerations acting on *Cassini* and on Dione, mainly the relativistic gravitational acceleration due to Saturn, its main satellites, the Sun, the other planets of the Solar System, the Moon, and Pluto. The masses and the states of the planets, the Moon, and Pluto were retrieved from the latest planetary ephemerides produced by JPL (DE438). The masses and the states of the Saturn satellites were retrieved from the latest satellite ephemerides produced by the JPL (SAT389). Both planetary and satellite ephemerides can be retrieved from ftp://ssd.jpl.nasa.gov/pub/eph/. Different versions of satellite ephemerides were also adopted, to test the stability of the solution.

Dione's gravity field was modeled as a linear combination of spherical harmonic functions:

$$U(r,\theta,\phi) = -\frac{GM}{r}\sum_{l=0}^{l_{\max}}\sum_{m=0}^{l}\left(\frac{R_{\text{ref}}}{r}\right)^l (C_{lm}\cos m\phi + S_{lm}\sin m\phi)P_{lm}(\cos\theta) \qquad (1)$$

where $C_{lm}$ and $S_{lm}$ are the unnormalized degree-$l$ and order-$m$ dimensionless gravitational potential coefficients for a reference radius $R_{\text{ref}}$ = 560 km, $P_{lm}$ are the unnormalized associated Legendre functions (e.g., Wieczorek, 2015), $l_{\max}$ is the maximum degree considered in the expansion, $GM$ is the product of the gravitational constant and Dione's mass (see Table 2), and the potential is evaluated here at radial position $r$, co-latitude $\theta$, and longitude $\phi$. In our discussion, we often represent the zonal terms in the expansion as $J_l = -C_{l0}$. In addition, the setup included the accelerations due to the extended gravity field of Saturn, modeled using the even zonal harmonics $J_2$-$J_{10}$ and $J_3$, as provided by the reference satellite ephemerides. For consistency, the corresponding rotational model of Saturn was adopted.

In estimating Dione's gravity coefficients (eq. 1), the minimum field capable of fitting the Doppler data to the noise level was a full degree 2. Higher degree and order fields were also estimated, to assess the stability of the solution. The time-variable gravity field of Dione caused by eccentricity tides was neglected, due to the very low orbital eccentricity and the limited data coverage. Considering only the encounters with coverage at C/A, for a $k_2$ tidal Love number of 0.5, the expected variation of $J_2$ and $C_{22}$ due to eccentricity tides is about 0.1% and 0.3%, respectively. These values are below the sensitivity of the measurements by almost a factor 2. Nevertheless, the $k_2$ tidal Love number of Dione was also estimated, as a stability test.

Regarding the rotational model of Dione, we adopted a dynamically defined, perfectly synchronous frame, which points always to the empty focus of the orbit (Murray and Dermott, 1999). In addition, to assess the stability of the solution, we alternatively assumed the rotational models suggested by IAU (see Section 2.5).

The dynamical model of *Cassini* included also the main non-gravitational accelerations: the Solar Radiation Pressure (SRP), and the thermal recoil pressure due to the anisotropic thermal emission, mainly caused by the three on-board Radioisotope Thermoelectric Generators (RTG). For both the accelerations, the models adopted by the *Cassini* navigation team were implemented (Di Benedetto, 2011). The accelerations due to the albedo and infrared thermal emission of Dione were neglected, being at least 1 order of magnitude smaller than the SRP (Di Benedetto, 2011).

Finally, during D4, two attitude maneuvers, executed with thrusters, were performed about 12h and 4h before closest approach. These maneuvers were modeled as impulsive changes in the spacecraft velocity vector, starting from the values reconstructed by the *Cassini* navigation team.

## 2.3. Data Selection and Calibration

The observable used in the estimation procedure was the range-rate, derived from the Doppler shift of a highly stable microwave carrier transmitted between *Cassini* and the ground antennas of NASA's Deep Space Network (DSN). The range observables, derived from the round-trip light time of a modulated code, were not used, because the information content provided is mainly related to planetary and satellite ephemerides, while the sensitivity to gravity field is limited.

The count time of Doppler data was chosen as a trade-off between the sensitivity to gravity spherical harmonics and numerical considerations. In fact, at the pericenter $r_p$ of a close flyby, the spatial scale $\Delta l$ of a spherical harmonic of degree $l$ is (Milani & Gronchi, 2010):

$$\Delta l = \frac{\pi}{l}r_p \qquad (2)$$

The time interval corresponding to the spatial scale at pericenter is obtained by dividing by the relative velocity:

$$\Delta t = \frac{\Delta l}{v_p} = \frac{\pi}{l} \frac{r_p}{v_p} \qquad (3)$$

The obtained time interval represents the theoretical maximum sampling time to correctly reconstruct the gravity field of degree *l*. Among all the *Cassini* flybys of Dione, the smallest time interval associated with the degree 2 field is 120 s, obtained during D3. This value represents a lower bound, because the orbit is hyperbolic. Therefore, Doppler data were integrated over 60 s, sufficiently smaller than the minimum time interval to be sensitive to the low degree gravity field and sufficiently large to avoid numerical noise issues (Zannoni & Tortora, 2013).

During the encounters, Doppler data at X (8.4 GHz) and Ka band (32.5 GHz) were acquired by the antennas of NASA's DSN, phase coherent to a common X-band (7.2 GHz) uplink. The analysis also used tracking data from standard navigation passes, covering about 5 days around the closest approach. This marginally improves the uncertainties in the gravity field estimation because of the improved reconstruction of the orbits of *Cassini* and Dione during the encounter. When available, X/Ka Doppler data were preferred over the standard X/X measurements, because they are less affected by the dispersive sources of noise, like solar plasma and Earth's ionosphere. When two-way Doppler data were not available, three-way data were also used. However, a bias on three-way data, constant per tracking pass, was estimated, accounting for a possible offset between the clocks of different DSN complexes. The additional path delay due to the Earth's troposphere was corrected using the standard GPS-based calibrations or, when available, the advanced calibrations based on Water Vapor Radiometers (Bar-Sever et al., 2007).

The data were analyzed using JPL's orbit determination program MONTE (Evans et al., 2018), currently used for the operations of several NASA deep space missions and for past radio science data analysis (e.g. Iess et al., 2018; Iess et al. 2019). The mathematical formulation of MONTE is described in detail in (Moyer, 1971) and (Moyer, 2000). Data were weighted using the observed RMS of the residuals, constant for each pass. Data acquired below 15 degrees of elevation, as viewed from the ground station, were discarded because of possible residual calibration errors of the Earth's troposphere and ionosphere, including data around the closest approach of D4 (between C/A - 1 h and C/A + 10 min).

## 2.4. Estimation

The data analysis was carried out using a multi-arc approach, in which radiometric data obtained during the different encounters are analyzed together to produce a single solution of a set of "global" parameters, which do not vary among the arcs (Milani & Gronchi, 2010). The multi-arc approach has been successfully applied to the analysis of radio science data of several deep space missions (Iess et al., 2012c; Iess et al., 2014; Modenini & Tortora, 2014; Tortora et al., 2016; Zannoni et al., 2018; Durante et al., 2019; Serra et al., 2019, Gomez Casajus et al., 2020). The parameters were estimated using a weighted least-squares batch filter, which determines corrections to an a-priori dynamical model to minimize the difference between the real and the simulated measurements (Bierman, 1977).

The set of global parameters includes the gravitational parameter (*GM*) of Dione, its full degree-2 gravity field, and *Cassini*'s RTG acceleration at a reference epoch. The a priori uncertainties of *Cassini*'s RTG acceleration, Dione's *GM*, $J_2$, and $C_{22}$ were chosen to avoid constraining the solution. No hydrostatic equilibrium constraint between $J_2$ and $C_{22}$ was imposed. For $C_{21}$, $S_{21}$, and $S_{22}$ we used a different strategy. From MacCullagh's theorem, these gravity coefficients are related to a misalignment between the adopted Dione-fixed frame and its principal axes of inertia. Because the data are not sufficiently sensitive to the rotational state of Dione, the a priori uncertainties of $C_{21}$, $S_{21}$, and $S_{22}$ were set to a value corresponding to a rotation of about 1°. Larger values were also used, up to a misalignment of 20°, to assess the stability of the solution.

In addition, a set of "local" parameters, affecting only a single encounter, was estimated. For each encounter, they include the initial state of *Cassini* and Dione, a constant correction to the SRP acceleration, constant Doppler bias for the three-way passes, and the impulsive ΔV due to the maneuvers executed during D4. The a priori uncertainties for *Cassini*'s position and velocity were 10 km and 0.1 m/s, respectively.

## 2.5. Results

The estimated gravity field coefficients of Dione are reported in Table 2, while Figure 2 shows the estimated values of $J_2$ and $C_{22}$ in the $C_{22}$-$J_2$ plane. Dione's quadrupole is dominated by $J_2$ and $C_{22}$, as expected by a satellite in synchronous rotation around its planet. However, the ratio $J_2/C_{22}$ is 4.102 ± 0.044, about 17-σ away from the ideal hydrostatic value of 10/3. Therefore, Dione's gravity field is significantly non-hydrostatic, meaning that the moment of inertia cannot be inferred directly from either $J_2$ or $C_{22}$ using the Radau-Darwin approximation—the interpretation requires a more sophisticated approach (see Section 3). The correlation coefficient between $J_2$ and $C_{22}$ is -0.5: although this value is not ideal (due to the limited coverage of the flybys), the important point is that the estimation of both parameters was not artificially constrained.

Table 2: Estimated values and 1-σ formal uncertainties of Dione's quadrupole gravity unnormalized coefficients. The adopted a priori values and uncertainties are also shown. The reference radius for the spherical harmonics is 560 km. The degree-1 terms are assumed zero, so that the origin of the Dione body-fixed frame is the moon's center of mass. The estimated value of the ratio $J_2/C_{22}$ and the correlation coefficient between $J_2$ and $C_{22}$ are also shown.

|  | Unit | A priori | Solution |
|---|---|---|---|
| *GM* | km³/s² | 73.116 ± 0.025 | 73.1118 ± 0.0025 |
| $J_2$ | x10⁶ | 1430 ± 500 | 1496 ± 11 |
| $C_{21}$ | x10⁶ | 0 ± 60 | 0.6 ± 6.8 |
| $S_{21}$ | x10⁶ | 0 ± 20 | 4 ± 20 |
| $C_{22}$ | x10⁶ | 365 ± 130 | 364.8 ± 1.8 |
| $S_{22}$ | x10⁶ | 0 ± 20 | -14.2 ± 1.9 |
| $J_2/C_{22}$ |  | 3.9 ± 1.9 | 4.102 ± 0.044 |
| corr $J_2$-$C_{22}$ |  | 0.0 | -0.50 |

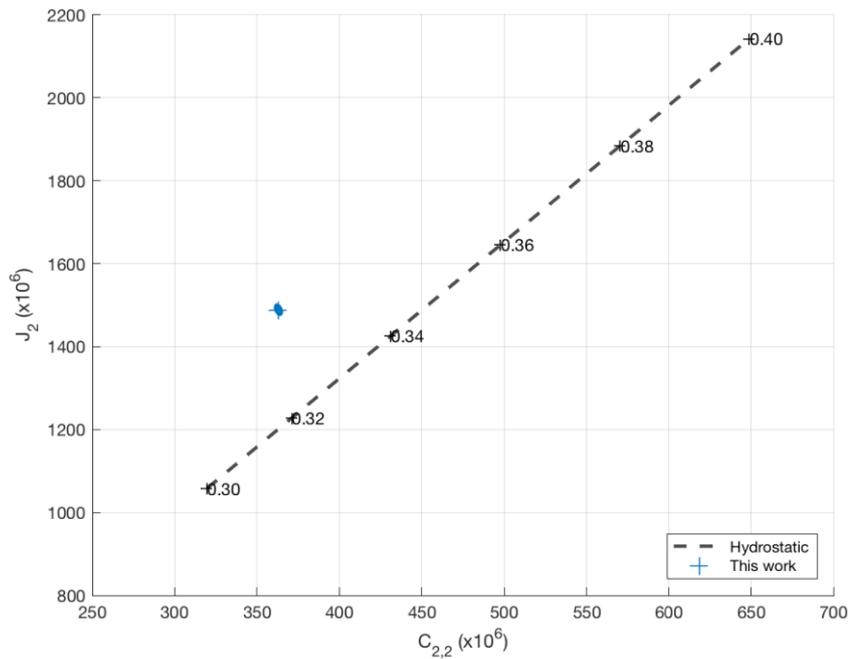

**Figure 2: Amplitude of latitudinal verses longitudinal variations in the gravitational field ($J_2$ vs $C_{22}$). Dashed line represents the expectation for a body in perfect hydrostatic equilibrium assuming various possible moments of inertia ranging from 0.3 to 0.4 (0.4 corresponds to a uniform body). The theoretical hydrostatic values are computed following the approach of (Tricarico, 2014).**

The estimated value of $C_{21}$ and $S_{21}$ are null within 1-$\sigma$. These values correspond to misalignments between the assumed spin axis and the maximum inertia axis of 0.02° ± 0.17° and -0.3° ± 1.5° around the y and x axes, respectively. However, the uncertainty of $S_{21}$ does not significantly improve with respect to the adopted a priori value, confirming a relatively low sensitivity of *Cassini*'s tracking data to the rotational state of Dione, in particular to rotations around its assumed prime meridian. $S_{22}$ is larger, being about 7-$\sigma$ away from zero. This corresponds to a misalignment between the principal axis of inertia and the prime meridian used in the analysis (pointing to the empty focus of the orbit of Dione around Saturn) of about 1.12° ± 0.15°.

The stability of the solution has been assessed by perturbing the adopted dynamical model, the data selection, and the estimation setup (such as the a priori covariances). Given the strategy to update the satellite ephemerides only locally, different sets of a priori ephemerides were also adopted. Since the nominal solution was obtained using all available data, different combinations of encounters were also tested, such as using only one encounter or removing one encounter from the dataset. Additionally, we considered different rotational models of Dione, in particular the ones suggested by IAU (Seidelmann et al., 2001; Archinal et al., 2018). In all cases, the estimated values were compatible with the reference solution within 1-$\sigma$, and the residuals were of very similar quality.

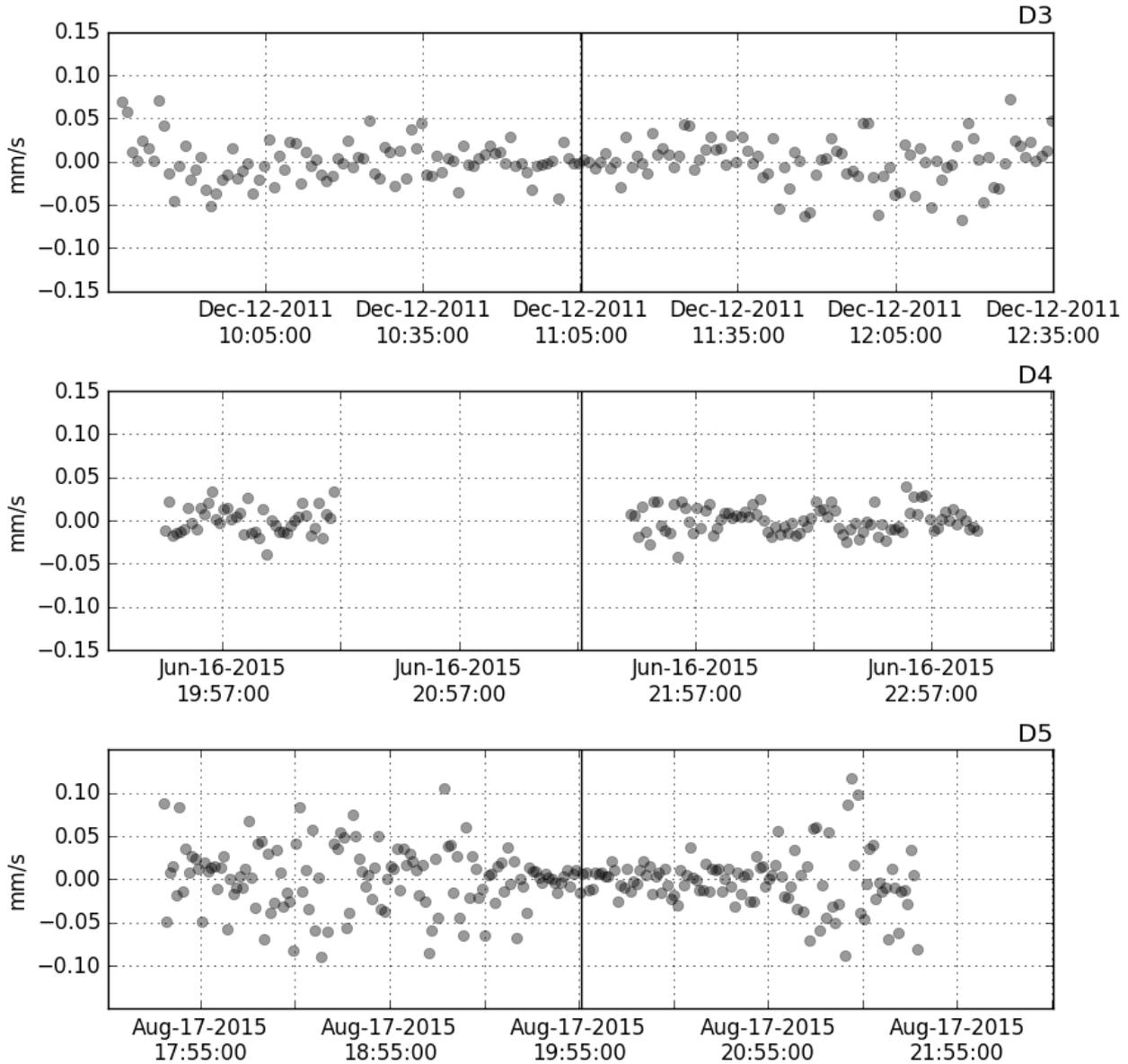

**Figure 3:** two-way range-rate residuals around the closest approach of D3, D4, and D5 encounters (vertical line), in mm/s. The quality of the residuals is good, which is an indication of a good orbital fit: no evident signatures are visible, the mean is approximately zero. The RMS is about 27 µm/s (D3), 21 µm/s (D4), and 36 µm/s (D5). As a comparison, the two-way range-rate signal due to the degree 2 field is about 20 mm/s (D3), 12 mm/s (D4), and 14 mm/s (D5). The closest approach of D4 is not covered because the elevation was lower than 15 degrees, as viewed from the ground station.

While a quadrupole gravity field is fully sufficient to fit the data to the noise level (Figure 3), the neglected higher degree components of the potential may introduce a bias in the estimation of $J_2$ and $C_{22}$. To test the robustness of the reference solution, we estimated also a gravity field of degree and order 4. However, given the number and the geometry of the *Cassini* flybys, an unconstrained estimation of the higher degrees is not possible. Thus, the a priori uncertainties on the normalized coefficients of degree *l* were set using the Kaula rule $K/l^2$ (Kaula, 1963). This empirical law can successfully describe the gravity power spectrum of the rocky planets, the Moon, and Vesta (Ermakov et al., 2018). A good agreement was also found for Titan, even if the gravity field is available only up to degree 5 (Durante et al., 2019). As of today, there are no geophysical arguments or empirical evidence to justify its applicability to the mid-sized icy moons of Saturn. However, even increasing the coefficient $K$ up to a very large value of $10^{-3}$, the quadrupole remains compatible within 1-$\sigma$ with the reference solution, confirming its stability.

The tides raised by Saturn produce a time variable component of the gravity field, which can be modeled using the $k_2$ tidal Love number. Even if the orbital eccentricity of Dione is small and the coverage provided by the *Cassini* encounters is limited, to test the robustness of the solution we tried also to estimate both the real and imaginary components of $k_2$. The estimated gravity coefficients remained compatible with the reference solution within 1-$\sigma$, while the estimated component of the Love number are Re($k_2$) = -0.01 ± 0.58 and Im($k_2$) = 0.04 ± 0.70, statistically equivalent to zero, confirming that the tidal response of Dione at the timescale of its orbital period is not observable using *Cassini*'s tracking data.

Constraining the ratio $J_2/C_{22}$ to the ideal hydrostatic equilibrium value of 10/3, the residuals show a large signature at the closest approach of D3 and D5 (Figure 4), confirming that *Cassini*'s data are not compatible with Dione being in hydrostatic equilibrium.

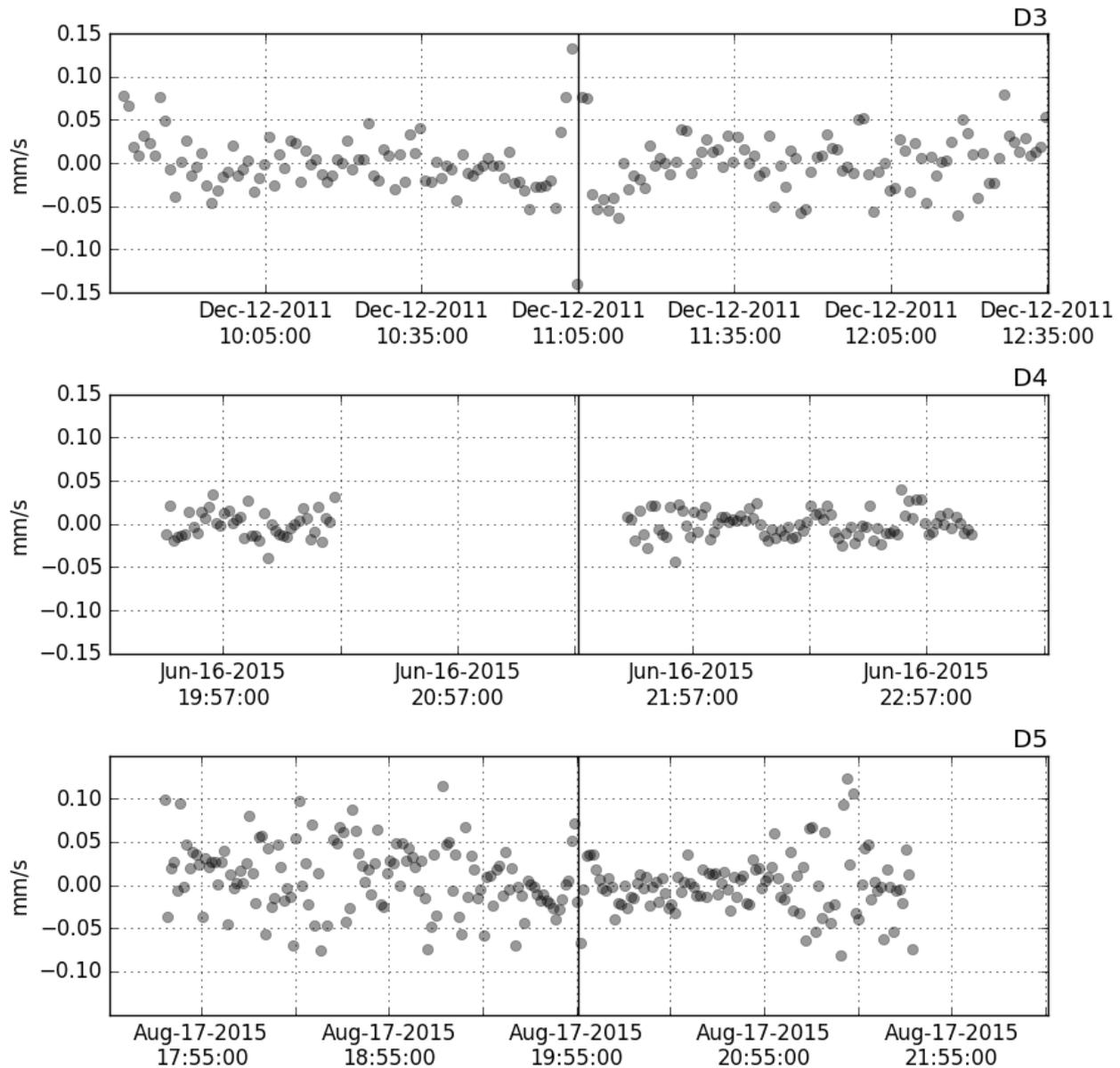

Figure 4: two-way range-rate residuals around the closest approach of D3, D4, and D5 encounters (vertical line), in mm/s, constraining the ratio $J_2/C_{22}$ to the ideal hydrostatic equilibrium value of 10/3. Strong signatures are visible around the closest approach of D3 and D5, not present in the unconstrained solution (Figure 3).

## 3. Interpretation

### 3.1. Basic Observations

Because their interiors are weak on long timescales, bodies as large as Dione are expected to have relaxed to hydrostatic equilibrium. That is, they should exhibit near spherical symmetry, with some small departures from symmetry arising due to centrifugal and tidal forces. The magnitude of these asymmetries (captured by the $J_2$ and $C_{22}$ gravity coefficients) is a function of the rotation rate, the mass and proximity of the parent body, and the body's internal radial density structure and therefore its moment of inertia. In general, one can use the Radau-Darwin relation (Darwin, 1899; Murray and Dermott, 1999) to compute the hypothetical hydrostatic $J_2$ and $C_{22}$ gravity coefficients for a range of possible moments of inertia, which can then be compared to the measured $J_2$ and $C_{22}$ to assess the degree to which the body's relatively stiff exterior supports a departure from the hydrostatic expectation. In our analysis, we employ the slightly more accurate approach of Tricarico (2014) in computing the expected hydrostatic figure (dashed black line in Figure 2, with slope ~3.307).

The statistically significant departure from hydrostatic equilibrium ($J_2/C_{22}$=4.102±0.044) makes it impossible to determine the precise moment of inertia (and hence radial density structure) directly from the measured gravitational field. It is, however, clear that Dione is far from the expectation for an undifferentiated hydrostatic body, for which the gravitational potential coefficients would be $J_2$=2127×10$^{-6}$ and $C_{22}$=649×10$^{-6}$ (the upper right terminus of the dashed black line in Figure 2, where the moment of inertia factor is 0.4). The measured gravitational potential is more consistent with some degree of differentiation and a moment of inertia factor of roughly 0.32-0.34 (Figure 2; more precise values are obtained below). For a simple two-layer interior model, consisting of a rocky core surrounded by an envelope of water ice (with density 930 kg/m$^3$), this moment of inertia range corresponds to core radii and densities of 380-430 km and 2100-2700 kg/m$^3$, respectively (Figure 5). Hence 56-66% of Dione's mass (31-46% of its volume) is in the rocky core. Assuming (non-hydrated) rock and ice densities of 3500 and 930 kg/m$^3$, respectively, yields a rock:ice ratio very close to 1:1 by mass (or about 1:4 by volume).

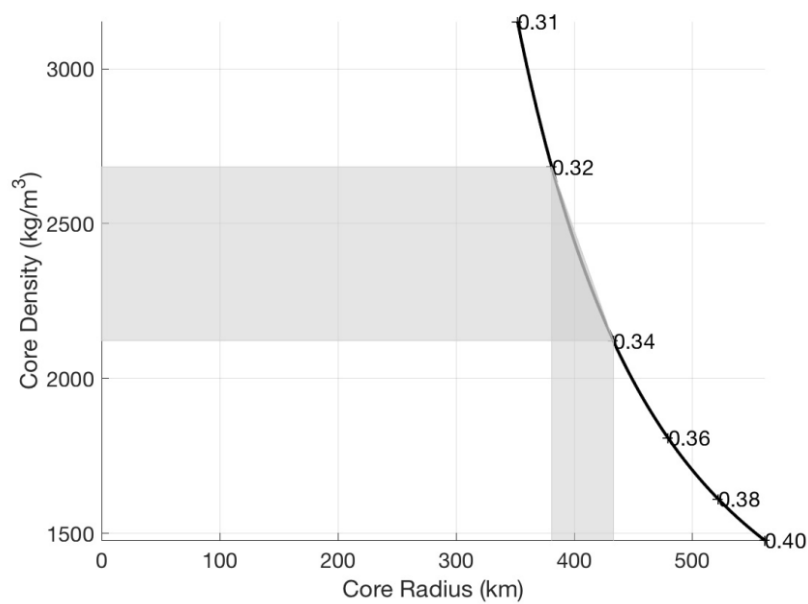

**Figure 5: Dione's core radius and density for a range of likely moments of inertia (0.32-0.34), assuming a simple two-layer model consisting of a rocky core surrounded by a water ice envelope with density 930 kg/m³. The total radius and bulk density are 564.1 km and 1478 kg/m³, respectively.**

To take the analysis further, we can combine the above gravitational field with a model of the shape. The radius and shape of Dione have been determined via analyses of limb profiles (e.g., Thomas et al., 2007; Thomas, 2010; Nimmo et al., 2011). Incorporating the latest limb profile observations (P. Thomas, personal communication), and repeating the analysis of Nimmo et al. (2011), we obtain an updated model for the degree-2 shape (Table 3, F. Nimmo, personal communication).

Table 3: The shape of Dione, based on analysis of *Cassini* ISS limb profiles, and represented here with unnormalized spherical harmonic expansion coefficients up to degree-2 (F. Nimmo, personal communication). In order to distinguish from the gravitational potential coefficients ($C_{lm}$), we have used the abbreviation $H_{lm}$ to represent the degree-*l* and order-*m* shape coefficients (uncertainties are 1-σ).

|  | Unit | Value |
| --- | --- | --- |
| $R$ | km | 561.4 ± 0.4 |
| $H_{20}$ | m | -1834 ± 134 |
| $H_{22}$ | m | 374 ± 13 |
| $H_{20}/H_{22}$ |  | -4.9 ± 0.4 |

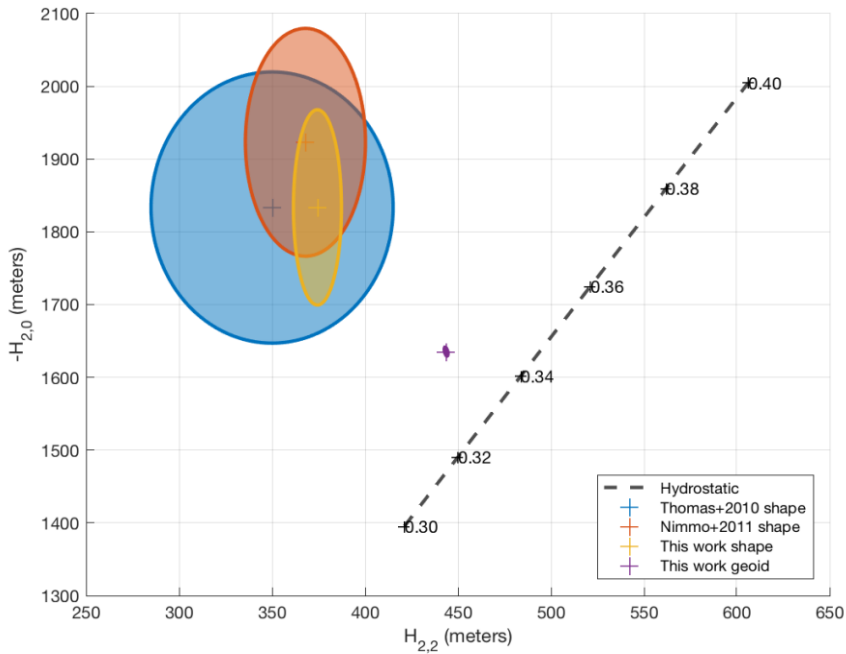

Figure 6: Amplitude of latitudinal verses longitudinal variations in the shape ($H_{20}$ vs $H_{22}$) for various shape models, including our new shape model (gold), in comparison with our computed geoid (purple). Dashed line represents the expectation for a body in perfect hydrostatic equilibrium assuming various possible moments of inertia ranging from 0.3 to 0.4 (0.4 corresponds to a uniform interior). The theoretical hydrostatic values are computed following the approach of Tricarico (2014). The negative value of $H_{20}$ is shown in order to maintain correspondence with plots of $J_2$ vs $C_{22}$.

Figure 6 shows our shape model along with previously published models (Thomas, 2010; Nimmo et al., 2011). Also shown in Figure 6 are the geoid (small purple ellipse) and the expectations for a perfectly hydrostatic Dione (dashed line), assuming various moments of inertia (i.e., corresponding to a range of possible radial density structures). The geoid coefficients are approximated by

$$N(\theta, \phi) = R_{\text{ref}} \sum_{l=0}^{\infty} \sum_{m=0}^{l} (C_{lm} \cos m\phi + S_{lm} \sin m\phi) P_{lm}(\cos\theta) - \frac{5\omega^2 R_{\text{ref}}^2}{6g} P_{20}(\cos\theta) + \frac{\omega^2 R_{\text{ref}}^2}{4g} P_{22}(\cos\theta) \cos 2\phi \quad (4)$$

where $\omega$ is Dione's rate of spin and $g$ = 0.232 m/s² is its surface gravity.

In spite of the considerable uncertainties in the shape model, it is clear from Figure 6 that, compared with the gravitational field (and the corresponding geoid), the measured shape exhibits a greater departure from the hydrostatic expectation, with the ratio between the main degree-2 coefficients being $H_{20}/H_{22}$=4.9±0.4 (recall that the corresponding ratio for the measured gravitational field is $J_2/C_{22}$=4.102±0.044). Because the $H_{22}$ component of the shape is smaller than the corresponding term for the geoid, the figure exhibits topographic highs on the leading and trailing faces when measured with respect to the observed geoid (Figure 7). This is unusual and the reason for it is not obvious. However, it is worth emphasizing that the most prominent topographic feature is the elevated region found near 60°E and just north of the equator (Figure 7b), and coinciding with parts of the bright chasmata found on the trailing hemisphere. A less pronounced, but broad topographic high also exists on the leading hemisphere. Although limb profile coverage is incomplete, it is sufficient to reveal these features clearly (Figure 8).

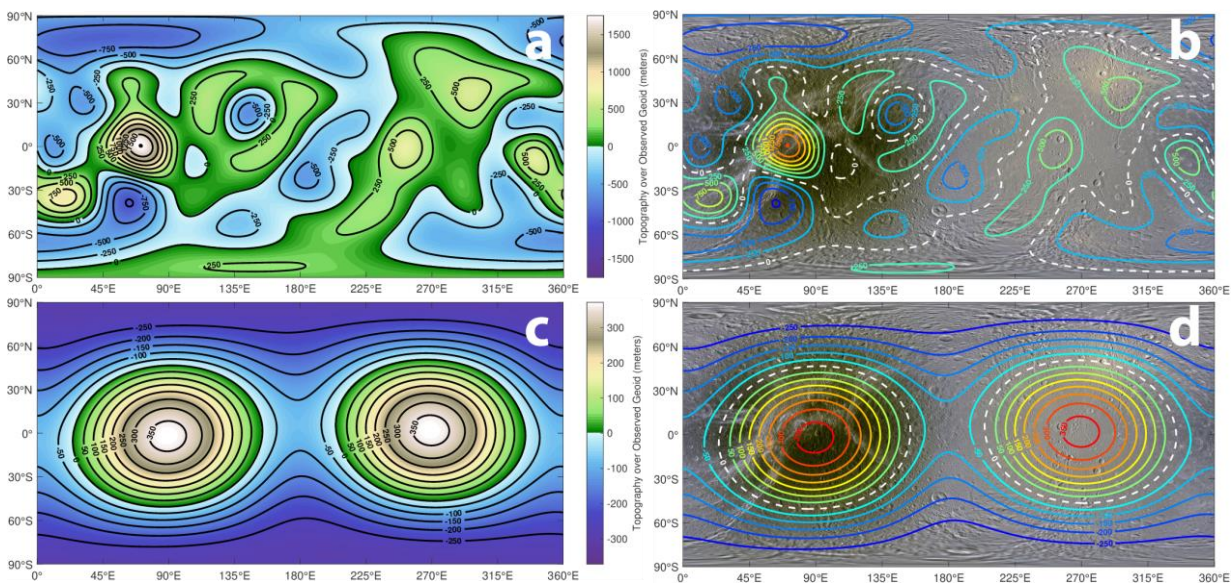

Figure 7: Dione's shape (Nimmo et al., 2011) with respect to the observed geoid (an equipotential surface), expanded to spherical harmonic degree 8 (a-b) and spherical harmonic degree 2 (c-d). Panels (b) and (d) show the topography as contours over a visible map of Dione produced by Paul Schenk (Lunar and Planetary Institute) from *Cassini* ISS data (NASA, JPL). Compared with the geoid, Dione's shape exhibits excess polar flattening and topographic highs on the leading and trailing faces (see also Figure 8).

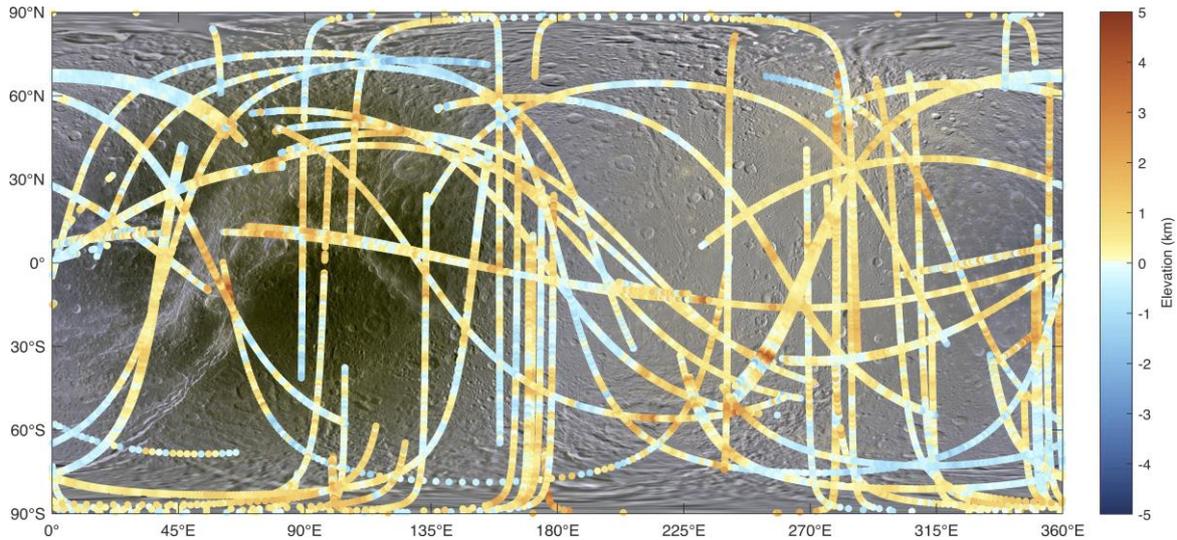

Figure 8: Limb profile elevation data (P. Thomas, personal communication) shown relative to the geoid, with warm and cool colors indicating high and low elevations, respectively (compare with our Figure 7 and with Figure 2d by Nimmo et al., 2011). The background image is a visible map of Dione produced by Paul Schenk (Lunar and Planetary Institute) from *Cassini* ISS data (NASA, JPL) and is intended to provide geographic context. Although limb profile coverage is sparse, topographic highs noticeably coincide with the chasmata on the trailing hemisphere (around 60-80°E) and, to a lesser extent, the leading hemisphere (around 270°E), whereas there is a topographic low near the anti-meridian (around 180°).

### 3.2. Isostatic Compensation and Interior Models

The existence of considerable non-hydrostatic topography indicates that the exterior of Dione has been cold and rigid enough to support the associated stresses since the formation of that topography, such that the figure has not completely relaxed to hydrostatic equilibrium. The fact that the corresponding non-hydrostatic gravity is smaller by comparison, however, is an indication that this non-hydrostatic topography is at least partly compensated (e.g., isostatically). That is, the topography's contribution to the gravitational field is partly offset by internal mass anomalies, likely resulting from lateral density variations and/or relief along internal density boundaries. The relationship between the non-hydrostatic gravity and the non-hydrostatic topography is a function of the degree and depth of this compensation, and therefore tells us about the shallower internal structure of Dione. The challenge, however, is to isolate these non-hydrostatic signals from the total observed shape and gravitational field, which are strongly affected by rotational and tidal deformation.

Following an approach developed for Enceladus (Hemingway et al., 2013; supplement of Iess et al., 2014), we consider Dione to be a mostly hydrostatic body (whose shape and gravitational field are dominated by rotational and tidal deformation), superimposed with some non-hydrostatic topography (of unspecified origin). That is,

$$H_{20}^{\text{obs}} = H_{20}^{\text{hyd}} + H_{20}^{\text{nh}}$$
$$H_{22}^{\text{obs}} = H_{22}^{\text{hyd}} + H_{22}^{\text{nh}}$$
(5)

For a moment of inertia factor of 0.33 (shown later to be representative of the best fitting interior models), it can be shown (see methods by Hemingway and Mittal, 2019) that the degree-2 figure and gravity break down into their hydrostatic and non-hydrostatic parts as shown in Table 4 and that the non-hydrostatic gravity expected if the topography were uncompensated is $C_{20}^{\text{nh}} \approx -188 \times 10^6$ and $C_{22}^{\text{nh}} \approx -64 \times 10^6$. The non-hydrostatic values in Table 4 being 63-65% of the uncompensated values is one measure of the degree of compensation.

Table 4: Decomposition of principal degree-2 terms into their hydrostatic and non-hydrostatic components.

|  | Unit | Observed | Hydrostatic | Non-hydrostatic |
|---|---|---|---|---|
| $H_{20}$ | m | -1834 | -1555 | -279 |
| $H_{22}$ | m | 374 | 469 | -95 |
| $C_{20}$ | x$10^6$ | -1464 | -1342 | -122 |
| $C_{22}$ | x$10^6$ | 365 | 406 | -40 |

We model Dione as consisting of a rocky (though not necessarily purely rock) core surrounded by an $H_2O$ envelope. To model the compensation of the surface topography, we consider the end member cases of Airy and Pratt isostatic equilibrium. For Airy compensation, in which the topography is supported by lateral variations in the thickness of the outer ice shell, this ice shell must be underlain by a higher density, lower viscosity material—most naturally a liquid water ocean. Hence, for our Airy models, we partition the $H_2O$ envelope into liquid and solid phases. For Pratt compensation, the topography is compensated by lateral density variations that persist through to some compensation depth. For these models, we partition the $H_2O$ envelope into an upper layer, in which there are lateral density variations, and a lower layer with some uniform density that is slightly greater than that of the variable density upper layer; both layers are assumed to be in the solid phase.

Following the approach of Hemingway and Mittal (2019), we construct a series of three-layer models with the exterior shape conforming to the observed values ($H_{20}^{\text{obs}}$, $H_{22}^{\text{obs}}$). The models are parameterized according to the mean thicknesses and densities of the two outer layers, yielding a four-dimensional parameter space. For each point in the parameter space, the mean radius and density of the innermost layer is constrained by the known total radius (561.4 km) and bulk density (1478 kg/m$^3$). For each interior model, we use the numerical approach of Tricarico (2014) to compute the hydrostatic terms ($H_{20}^{\text{hyd}}$, $H_{22}^{\text{hyd}}$). The remaining non-hydrostatic topography ($H_{20}^{\text{nh}}$, $H_{22}^{\text{nh}}$) is then assumed to be compensated isostatically. We compute the compensating basal topography (when assuming Airy compensation) or density variations (when assuming Pratt compensation) using the equal pressures isostasy approach of Hemingway and Matsuyama (2017). The shape of the innermost layer (the core) is assumed to conform to the hydrostatic expectation. This is a reasonable assumption if Dione's core is sufficiently unconsolidated to be weak on long timescales. Departures from a hydrostatic figure for the core are possible if the material is sufficiently strong (e.g., McKinnon, 2013; Tajeddine et al., 2014) but, as in previous work (Hemingway and Mittal, 2019), we regard a hydrostatic core as the simplest assumption since there is no reason to prefer non-hydrostatic core topography of any particular orientation. Allowing for a wider range of core shapes would widen the uncertainties in our internal structure models but would not bias our results in any particular direction. Finally, we compute the resulting gravitational field (Hemingway and Mittal, 2019, eq. 8), taking into account the finite amplitude effects (Wieczorek and Phillips, 1998), and compare the result with the observed gravitational field (Table 2). We use the misfit between the model and measured gravitational fields to construct a probability density function across the parameter space (see section 2.6 of Hemingway and Mittal, 2019), indicating which parameter values are most likely, given the observations. We carry out this exercise assuming either Airy (Figure 9) or Pratt (Figure 10) isostasy.

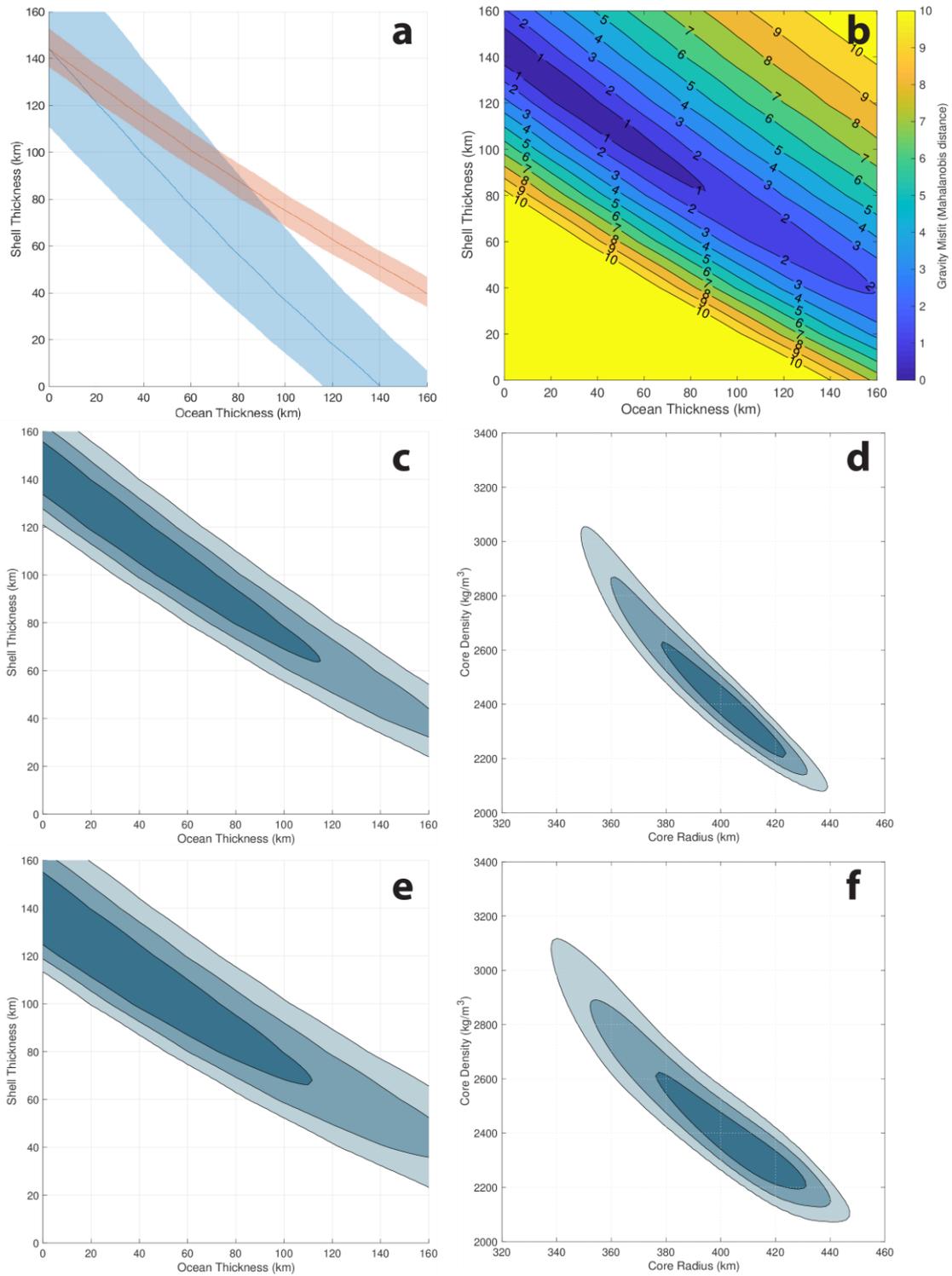

Figure 9: The range of likely interior structures for Dione assuming Airy isostasy. (a) Curves illustrate the range of parameters that satisfy the observed $J_2$ (blue) or $C_{22}$ (red) gravitational potential coefficients, with shaded bands indicating 1-$\sigma$ uncertainties. (b) Misfit between model and observed gravitational potential. (c) Probability density function, with 68% (dark), and 95% (intermediate), and 99.7% (pale) confidence contours showing the parts of the parameter space that best fit the observed gravitational potential. (d) Corresponding range of best fitting core radii and densities. Panels (a-d) assume ice and ocean densities of 925 kg/m³ and 1020 kg/m³, respectively. (e-f) Same as (c-d) except allowing for a wider range of possible ice and ocean densities (ice densities of 850-950 kg/m³; ocean densities of 1000-1100 kg/m³).

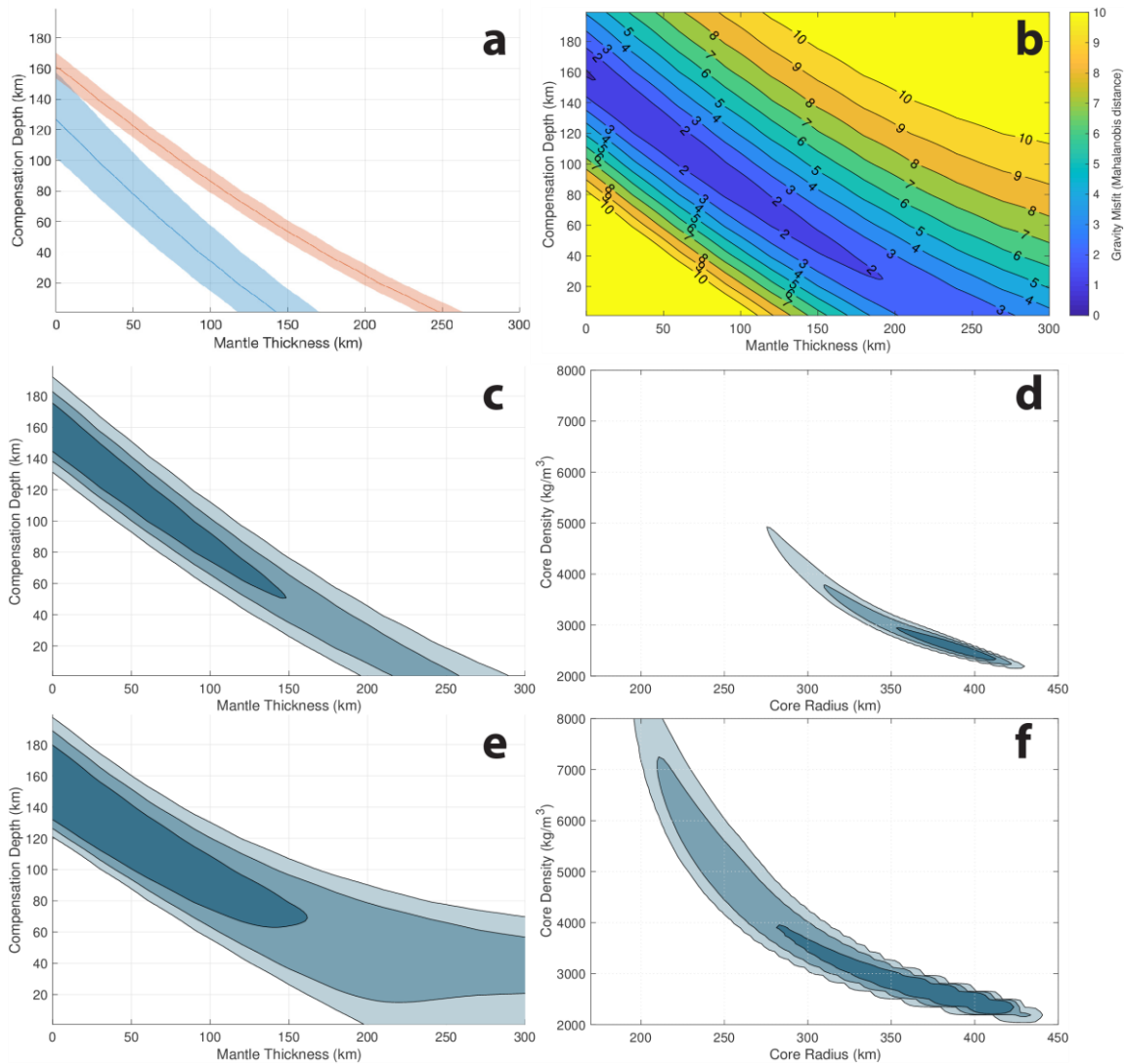

Figure 10: The range of likely interior structures for Dione assuming Pratt isostasy. (a) Curves illustrate the range of parameters that satisfy the observed $J_2$ (blue) or $C_{22}$ (red) gravitational potential coefficients, with shaded bands indicating 1-$\sigma$ uncertainties. (b) Misfit between model and observed gravitational potential. (c) Probability density function, with 68% (dark), 95% (intermediate), and 99.7% (pale) confidence contours showing the parts of the parameter space that best fit the observed gravitational potential. (d) Corresponding range of best fitting core radii and densities. Panels (a-d) assume ice and mantle densities of 925 kg/m$^3$ and 1030 kg/m$^3$, respectively. (e-f) Same as (c-d) except allowing for a wider range of possible densities (mean outer layer densities of 850-950 kg/m$^3$; mantle densities of 1000-1300 kg/m$^3$).

In the Airy isostasy model, the topography is supported by lateral thickness variations in an ice shell floating on a liquid water ocean. Assuming nominal ice shell and ocean densities of 925 kg/m$^3$ and 1020 kg/m$^3$, respectively, the best fitting interior models tend towards large shell thicknesses (as much as 140 km) and small ocean thicknesses. However, the data permit a considerable range of possibilities as smaller shell thicknesses can be traded off against larger ocean thicknesses (Figure 9c). Similar trade-offs exist between core radius and density, but the preferred range is approximately 400±25 km and 2400±200 kg/m$^3$ (Figure 9d). When a wider range of possible ice and ocean densities is considered, the range of best fitting parameters widens only slightly (Figure 9e-f); hence, the ice and ocean densities are not well constrained. An analysis by Beuthe et al. (2016), based on our preliminary results (Hemingway et al., 2016), delivered a similar range of best fitting interior models (compare their Figure 2b with our Figure 9c), in spite of their use of somewhat different methods for computing the equilibrium figures and the isostatically compensating basal topography.

In the Pratt model, not considered in previous work, the topography is compensated instead by lateral density variations in the outermost layer. Such density variations could be the result of differences in porosity within the outermost layer, for example. Assuming nominal mean densities of 925 kg/m$^3$ for the outermost layer, and 1030 kg/m$^3$ for the intermediate layer (which we call the mantle), the best fitting interior models correspond to a compensation depth of roughly 160 km, though again there is a range of possibilities as compensation depth can be traded to some degree against the thickness of the underlying mantle. The range of likely core radii and densities is very broad. The range of possibilities widens when we consider a broader range of mean layer densities. In all cases, however, the best fitting results correspond to compensation depths greater than 60 km. Confining the density anomalies to a shallower layer would require more pronounced density anomalies, leading to values of $J_2$ and $C_{22}$ that would exceed the observed value (i.e., as we move below the curves shown in Figure 10a).

### 3.3. Discussion

Because we do not evaluate the probabilities in absolute terms, the probability density functions in Figures 9 and 10 cannot be compared directly. Instead, only the misfits (Figures 9b and 10b) can be compared directly. Although both the Airy and Pratt end member scenarios permit solutions within the measured uncertainties (Figures 9a, 10a), the Airy scenario yields much smaller misfits and thus accommodates the data much better. Moreover, it may be difficult for the Pratt mechanism to dominate given that it requires the lateral density anomalies to persist to depths of several tens of kilometers. If the density anomalies are due to variations in porosity, they may not be able to reach such depths given the high overburden pressures (>2 MPa per 10 km of depth, reaching ~13 MPa at 60 km) and increasing temperatures with depth (~1 K/km assuming a 120 km thick conductive shell), both of which effects would tend to close those pores (Besserer et al., 2013). Dione's excess topography may therefore require some degree of Airy-type isostasy, meaning that the outer ice shell could be underlain by a higher density, lower viscosity layer—most straightforwardly interpreted as a subsurface liquid water ocean (Figure 11a).

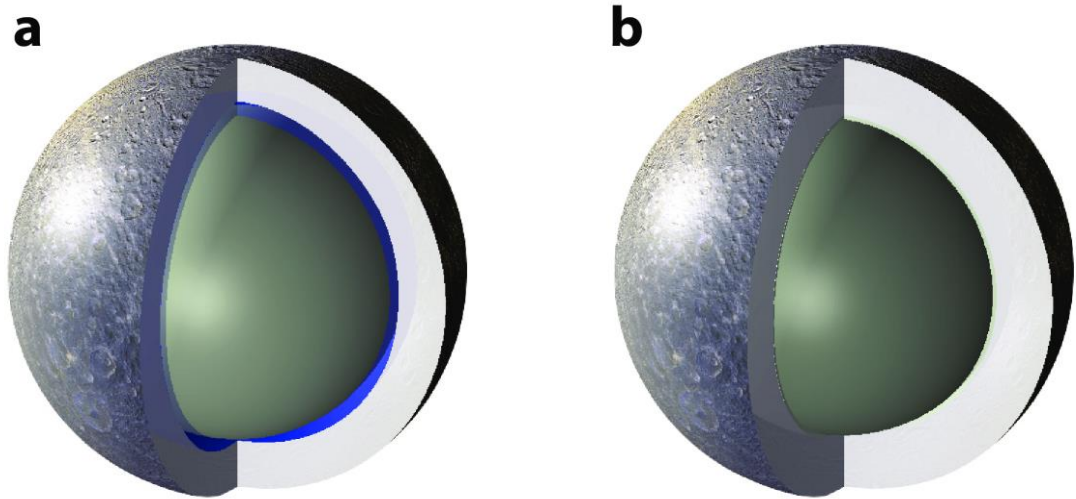

Figure 11: Depictions of the interior of Dione, to scale. (a) Interior when Airy compensation is assumed. The model matches the observed external shape and reproduces the observed gravitational field exactly. The model comprises an outer ice shell with a mean thickness of 120 km, a subsurface liquid water ocean with a mean thickness of 30 km, and a 411 km radius core, with a density of ~2300 kg/m³; the moment of inertia factor is ~0.332. (b) Interior when Pratt compensation is assumed. The best fitting Pratt model reproduces the observed gravitational field within the uncertainties. The model comprises a ~400 km radius core, with a density of ~2400 kg/m³, and an H$_2$O envelope with a mean thickness of ~160 km; lateral density anomalies persist throughout this layer (but see text); the moment of inertia factor is ~0.327. Surface texture is a global visual map of Dione produced by Paul Schenk (Lunar and Planetary Institute) from *Cassini* ISS data (NASA, JPL).

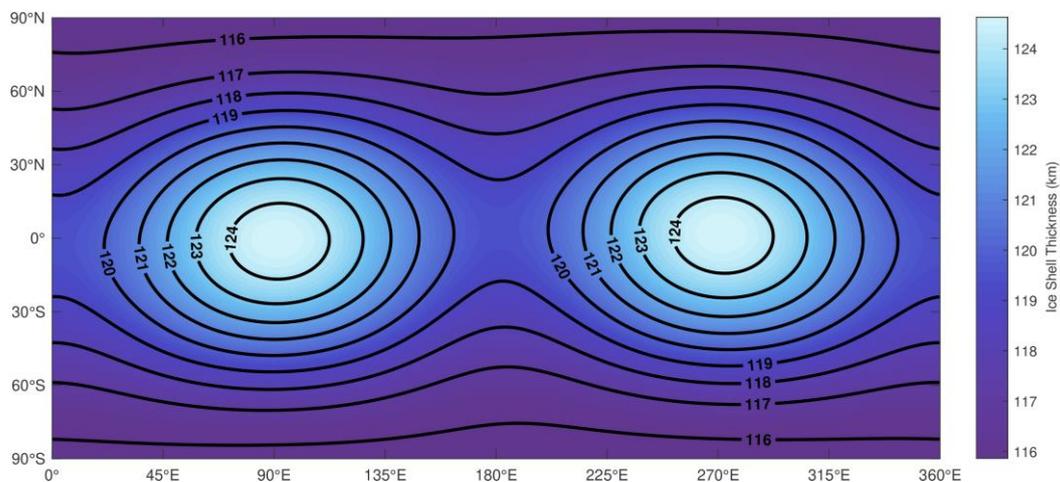

Figure 12: Dione's ice shell structure assuming Airy isostasy (where the effects of topography beyond spherical harmonic degree 2 are excluded). The thickest parts of the ice shell correspond with the regions of highest-standing non-hydrostatic topography, roughly centered on the leading and trailing faces.

If Dione does indeed harbor an internal liquid water ocean (Figure 11a), then the approximately known temperature of the ice/ocean interface (i.e., near 270 K) allows us to place a lower bound on the rate of heat loss. In the most conservative case, where the entire ice shell behaves conductively, with conductivity going as $T^{-1}$, the temperature structure can be described by

$$T(r) = T_s \left(\frac{T_b}{T_s}\right)^{\left(\frac{R}{r}-1\right)\left(\frac{R}{d}-1\right)} \tag{6}$$

where $T_b$ and $T_s$ are the basal and surface temperatures, respectively, *d* is the mean ice shell thickness, and *R* is Dione's mean radius. The heat flux at the surface is given by

$$F = \frac{c}{d}\ln\left(\frac{T_b}{T_s}\right)\left(1-\frac{d}{R}\right) \tag{7}$$

where *c* is an empirical constant capturing the temperature dependence of the thermal conductivity, $k = c/T$, where we take *c*=651 W/m (Petrenko and Whitworth, 1999, p.43). Taking the mean surface temperature to be $T_s = 87$K, and assuming an ice shell thickness of 120 km, the total conductive heat loss is approximately 19 GW (~4.8 mW/m$^2$). Although studies of elastic flexure (Hammond et al., 2013) and crater relaxation (White et al., 2017) have suggested episodes of even greater heat flows in Dione's past, such an intense rate of heat loss is difficult to sustain without the ocean freezing out. For Dione, assuming an ice shell thickness of 120 km, ocean freezing occurs at a rate of ~40 meters per million years per 1 GW of net heat loss, meaning that a 30-km thick ocean would freeze completely within ~40 Myrs if this 19 GW of heat loss were not balanced by internal heat production. Assuming a chondritic heating rate of $4.5 \times 10^{-12}$ W/kg (Spohn and Schubert, 2003) and Dione's rock mass of $\sim 5.5 \times 10^{20}$ kg (again assuming the chondritic rock mass density is 3500 kg/m$^3$), radiogenic heat production is ~2 GW. Given the recent finding that Saturn may be more dissipative than previously thought (Lainey et al., 2012, 2016, 2017, 2020; Fuller et al., 2016), tidal heating is potentially large enough to balance the heat budget, but more work is required to determine whether, how, and where tidal heating within Dione's interior could be sufficient to prevent an ocean from freezing completely. Maintaining an internal liquid water ocean would be even more difficult if the ice shell were undergoing solid state convection—a possibility that cannot be ruled out given that, unless very high effective viscosities are assumed, such a thick ice shell yields large Ralyeigh numbers (e.g., Kamata, 2018).

If the ice shell is entirely conductive, then from (6), and again assuming $T_s = 87$K and an ice shell thickness of 120 km, we expect the upper 57-83 km to be cold enough (i.e., below 140–180 K) to behave elastically, even on long timescales (e.g., Nimmo et al., 2002). On the other hand, our Airy isostasy model assumes effectively no elastic support for the long wavelength (degree-2) topography. In principle, including the effect of elastic support would shift our shell thickness estimates to smaller values if the topography were generated at the surface, or to larger values if the topography were generated due to heterogeneous freezing and melting at the ice/ocean interface (Hemingway and Mittal, 2019). Our model for complete Airy compensation and the implied present-day temperature structure are thus mutually consistent only if the topographic anomalies formed when the effective elastic layer thickness was negligible (i.e., at times of higher heat flow; Hammond et al., 2013; White et al., 2017) or if the elastic layer is sufficiently fractured to make long wavelength elastic support ineffective. For example, following McKinnon (2013) and Beuthe et al. (2016), we can estimate the tangential lithospheric stresses resulting from the deflection required by the non-hydrostatic topography, assuming a shear modulus of 3.5 GPa. For the case of bottom loading, which would imply ~300 meters of deflection to produce the degree-2 topography (Figure 7c), we estimate stresses of ~30 MPa—far in excess of the ~1–2 MPa failure limit for intact ice (Schulson and Duval, 2009). In the case of top loading, the lithospheric deflection required to satisfy the gravity observations depends on the compensation depth: complete Airy compensation at a depth of 120 km produces an effect similar to that of partial compensation at a shallower depth. However, given that the non-hydrostatic gravity is ~0.65 of the uncompensated value, it can be shown that, even in the limit of very shallow compensation, the required deflection is >1 km, leading to stresses of more than 100 MPa. Hence, regardless of the loading history and temperature structure, elastic support is unlikely to be effective for the observed degree-2 topography.

Since the high-standing non-hydrostatic topography is found on the leading and trailing faces, Airy isostasy would imply that these are the thickest parts of the ice shell (Figure 12). Why the ice shell should be thickest on the leading and trailing faces, however, is not clear. Heterogeneous tidal dissipation within the ice shell should lead to lateral shell thickness variations (Ojakangas and Stevenson, 1989; Beuthe, 2013; Hemingway and Mittal, 2019), but the thickest parts of the shell are expected to be at the prime- and anti-meridians, where tidal dissipation is weakest, not on the leading and trailing faces. This 90-degree longitudinal shift in the locations of the thickest parts of the shell might be suggestive of non-synchronous rotation or an ice shell reorientation event, perhaps contributing to the mismatch between the theoretical and observed apex-antapex asymmetry in cratering (Leliwa-Kopystynski et al., 2012). It is also worth noting that the highest standing topography, measured relative to the geoid, coincides with the prominent series of chasmata found on the trailing hemisphere (Figure 7b; Figure 8). These features may be expected to exhibit lower densities and thus may be partly compensated in the Pratt sense. However, it is not clear whether such density anomalies could persist to the depths (>60 km) required for this to significantly account for the observed compensation.

Finally, we note that our best fitting core density of 2300–2400 kg/m$^3$ is very close to the inferred core density for Enceladus at 2340–2410 kg/m$^3$ (Hemingway and Mittal, 2019) in spite of the two bodies having somewhat different bulk densities (1478 kg/m$^3$ for Dione, 1609 kg/m$^3$ for Enceladus). Similarly, for Titan, whose bulk density is 1881 kg/m$^3$, the core density may be as low as ~2500 kg/m$^3$, if one assumes a hydrostatic interior and a density of <1000 kg/m$^3$ for the H$_2$O envelope (Durante et al., 2019). Although we do not have a ready explanation for the similarities in these core densities, we bring attention to this observation because, if it is not a mere coincidence, it could have implications for the formation processes of the Saturnian satellites and icy moons in general.

## 4. Conclusions

We presented an estimation of the gravity field of Dione, obtained by analyzing the *Cassini* Doppler tracking data acquired during five close flybys of the moon. A full degree 2 field was sufficient to fit the data to the noise level. The estimated values of the principal quadrupole terms, $J_2 \times 10^6 = 1496 \pm 11$ and $C_{22} \times 10^6 = 364.8 \pm 1.8$ (unnormalized coefficients, 1-$\sigma$ uncertainty), and their ratio, $J_2/C_{22} = 4.102 \pm 0.044$, indicate a significant departure from the expectation for a body that has relaxed to hydrostatic equilibrium. The departure from hydrostatic equilibrium means that the moment of inertia cannot be inferred directly from the gravitational field, but a combined analysis of gravity and topography suggests a substantial degree of differentiation, with a moment of inertia factor of approximately 0.33. The analysis further demonstrates that the high-standing topography is largely compensated by some combination of lateral density anomalies and the deflection of internal density interfaces (i.e., Pratt and Airy isostatic support, respectively), the latter mechanism being consistent with the presence of an internal liquid water ocean. The Airy model better accommodates the observational constraints, but an internal liquid water ocean would imply rapid heat loss requiring some 17 GW of tidal heating to keep the ocean from freezing.

Further insights into Dione's interior may come from an improvement of the topography or gravity models, and from the development of more sophisticated geophysical and geochemical models. In particular, reconstructing a coherent trajectory of Dione during the entire timespan of the *Cassini* mission has the potential to decrease the uncertainty on the gravity coefficients by approximately a factor of 2, though we note that the uncertainties in our interior models are rather dominated by uncertainties in the shape models derived from limb profile analyses. After the spectacular end of the *Cassini* mission in 2017, there are no currently planned missions to study Dione or the other mid-sized icy moons of Saturn. However, a future mission dedicated to a comprehensive characterization of these bodies could shed further light on their interiors, leading to an improved understanding of the formation and evolution of the Saturn system, and of icy moons in general.

## Acknowledgements

The authors are grateful to Luciano Iess, John W. Armstrong, and Sami W. Asmar of the *Cassini* Radio Science Team, for the useful discussions and suggestions regarding the procedures for *Cassini* data analysis. We thank Peter Thomas for providing updated limb profiles and Francis Nimmo for using these to generate the updated degree-2 shape model used in our analysis. M.Z., L.G.C., and P.T. acknowledge support from the Italian Space Agency through the Agreement 2017-10-H.O in the context of the NASA/ESA/ASI mission, and Caltech and the Jet Propulsion Laboratory for granting the University of Bologna a license to an executable version of MONTE Project Edition S/W. D.H. was funded by the Miller Institute for Basic Research in Science at the University of California Berkeley, and by the Carnegie Institution for Science in Washington, DC. We thank the *Cassini* Project, A. Anabtawi, and the JPL Planetary Radar and Radio Science Group, the NASA/JPL Deep Space Network and their operations personnel involved in the acquisition of Doppler data analyzed here.

## Data Availability

The Doppler data and ancillary information used in this analysis are archived in NASA's Planetary Data System. https://pds.nasa.gov